\magnification=\magstep1
\baselineskip=20pt
\hsize=16.5truecm
\vsize=23truecm
\font\ftit=cmbx10 scaled \magstep 1
\font\fnom=cmr10

\def\sseccion#1 #2  {\goodbreak \vskip1.5truecm plus2truecm
minus1truecm {\bf  
\leftline {#1}\leftline {#2}} \vskip1truecm
plus.5truecm \nobreak} 
\raggedbottom
\def\ii{\'\i}

\fnom

\newbox\Ancha
\def\gros#1{{\setbox\Ancha=\hbox{$#1$}%
\kern-.025em\copy\Ancha\kern-\wd\Ancha
\kern.05em\copy\Ancha\kern-\wd\Ancha
\kern-.025em\raise.0433em\box\Ancha}}



\baselineskip=16pt

\centerline{\ftit A General     Algebraic Model for }

\centerline{\ftit  Molecular Vibrational Spectroscopy} 

\

\

\centerline{\bf  A. Frank$^{1,2}$, R. Lemus$^1$, R. Bijker$^1$, F.
P\'erez-Bernal$^3$ and J.M. Arias$^3$ } 

\

\centerline{$^1$ {\it Instituto de Ciencias Nucleares, UNAM } }
\centerline{\it Apdo. Postal 70-543, Circuito Exterior, C.U.} 
\centerline{\it 04510 M\'exico, D.F., M\'exico.} 

\

\centerline{$^2$ {\it Instituto de F\ii sica, Laboratorio de
Cuernavaca, UNAM} }
\centerline{\it Apdo. Postal 139-B,  Cuernavaca, Morelos, M\'exico} 

\

\centerline{$^3$ {\it Departamento de F\ii sica At\'omica, Molecular
y Nuclear} }
\centerline{\it Facultad de F\ii sica, Universidad de Sevilla} 
\centerline{\it Apdo. 1065, 41080 Sevilla, Espa\~na}

\

\

\

\baselineskip=20pt

\centerline{\ftit ABSTRACT} 

\

We introduce the Anharmonic Oscillator Symmetry Model  to describe
vibrational excitations in 
molecular systems exhibiting high degree of symmetry.  A systematic
procedure 
is proposed to establish the relation between the algebraic and
configuration space formulations, leading to new interactions in the
algebraic model.  This approach  incorporates the
full power of group theoretical techniques and  provides reliable
spectroscopic predictions.  We illustrate the  method for the case of
${\cal D}_{3h}$-triatomic molecules.

\
\vfill
\eject

\noindent
{\ftit 1. Introduction} 

\

Spectroscopic techniques represent one of the most important tools in
modern chemical analysis.$^{1)}$  In particular, the molecular
vibrational degrees of freedom are studied by means of Infrared and
Raman spectroscopy.$^{2)}$  It is necessary, however, to rely on
theoretical models in order to interpret the data, which in turn
refines the models in a feedback cycle. 

The study of molecular vibrational excitations is carried out by
taking into account different degrees of approximations and
theoretical assumptions.  The simplest way to study the molecular energy
spectra is by means of a Dunham expansion.$^{2)}$ This method,
however, does not provide wave functions, and consequently does not
allow the calculation of  physical properties such as
transition intensities.  On the other  hand  there are {\it ab
initio} calculations, where an exact solution of the Schr\"odinger
equation is attempted.  In practice, the molecular Hamiltonian is
usually parametrized as a function of internal coordinates and 
the potential  modeled in terms of force field constants,$^{3)}$
which are determined either through calculations involving the
molecular electronic states  for several configurations$^{4)}$ or
empirically, by the fitting of experimental data.$^{5)}$  While for
diatomic and triatomic molecules very accurate information on force
field constants is available,$^{6)}$ this is not the case for
polyatomic molecules, due to the large size of their configuration
spaces.  It is thus important to develope  alternative calculational
methods in order to describe complex molecules for which {\it ab
initio} calculations are not feasible. 

Algebraic models attempt to provide such  alternative techniques.  In
1981 an  algebraic approach was 
proposed to describe the roto-vibrational structure of diatomic
molecules,$^{7)}$   subsequently  extended to  linear
tri- and 
four- atomic molecules$^{8)}$ and certain  non-linear  triatomic
molecules.$^{9)}$  Although these  results were encouraging, the
model  could not  in practice 
be extended to  polyatomic molecules, for which  it is
necessary to incorporate the  underlying 
discrete symmetries.  This difficulty can be surmounted by treating
the vibrational degrees of freedom separately from the rotations.  In
1984 Van Roosmalen {\it et al} proposed  a U(2)-based model
to describe the stretching vibrational modes in ABA
molecules,$^{10)}$ later 
extended to describe the stretching vibrations of polyatomic
molecules such as octahedral and benzene like molecules.$^{11)}$
Recently the bending modes have also been incorporated  to the
framework, which was then applied to 
 describe ${\cal
C}_{2v}$-triatomic molecules$^{12)}$ and the lower excitations of
tetrahedral molecules,$^{13)}$ using  
a scheme which combines Lie-algebraic and point group methods.  
 In a different approach, it has also been 
 suggested  to use a $U(k+1)$ model for the $k =
3n-3$ rotational and vibrational degrees of freedom of a $n$-atomic
molecule.  This model has the advantage that it  incorporates
all rotations and vibrations and takes into  account
 the relevant point group symmetry,$^{14)}$  but  for larger
molecules the  number of possible interactions and the 
size of the Hamiltonian matrices increase very
rapidly, making  it impractical to apply.

The  algebraic formulations have no doubt proved useful, but several
problems remained, the most important of which is the 
absence of a clear connection to configuration space traditional
methods, which in turn 
makes their significance difficult to gauge.  A  
related problem  is the lack of a systematic
procedure to construct the physically meaningful interactions in the
algebraic space.  In this paper we address both these issues and
introduce a general model for the analysis of molecular vibrational
spectra, which we call the Anharmonic Oscillator Symmetry Model
(AOSM).  We shall show that it is possible to construct algebraic
operators with a well defined physical interpretation and in
particular the interactions which are of
special relevance for the description of the degenerate modes
present in systems exhibiting high degree of symmetry.  These
 are derived in a procedure that  takes full advantage of
the discrete  symmetry of the molecule and that provides  all
possible terms in a 
systematic fashion.  As a bonus, a clear-cut connection is
established between the algebraic scheme and the traditional
analyses based on internal coordinates, which  correspond to   the
harmonic limit of the model. 

As a test for this approach 
  we apply the AOSM  to three
${\cal D}_{3h}$-triatomic molecular systems, namely Na$^+_3$, Be$_3$ and
H$^+_3$.  We have chosen this set of molecules because they cover a
wide range of behaviors.   Whereas Na$^+_3$ is very harmonic, H$^+_3$
 displays a strongly anharmonic spectrum, while the Be$_3$ cluster
has an intermediate behavior.  Since these small molecules can
be very well described by means of {\it ab initio}
calculations,$^{15,16)}$ we again emphasize the aim of this work.
We  shall establish  an exact  correspondence
between configuration space and algebraic interactions  in 
the harmonic limit of the U(2) algebra.  
  This general procedure not only allows to  derive the
interactions in the AOSM  from  interactions in
configuration space, but can also  be applied to cases for which no
configuration space interactions are available.  
  The ${\cal D}_{3h}$-triatomic molecules constitute the
simplest systems where degenerate modes appear and where the new
interactions in the  model become significant.  The application
of these techniques to more complex systems, such as tetrahedral
molecules,  
 is  presented elsewhere.$^{17)}$ 

In the next section the structure of the  model is presented.
Section 3 is devoted to the construction of the symmetry adapted
normal basis, which is the most appropriate one to carry out the
diagonalizations.  In Section 4 we propose a new set of interactions
which have physical interest and suggest the need to construct
operators associated to the $E$ mode.  In Section 5 we describe the
general procedure to derive the algebraic interactions from those
appearing in configuration space and in Section 6 we introduce the
AOSM in order to derive all 
algebraic interactions from symmetry considerations. In Section 7 we
apply the model to 
H$^+_3$, Be$_3$ and  Na$^+_3$, while in Section 8 we present our
conclusions and discuss some future developments of the model.

\vfill
\eject

\noindent
{\ftit 2.  Algebraic Model} 

\

The   model  exploits  the isomorphism of the $U(2)$ Lie
algebra and the one dimensional Morse oscillator 
$$
{\cal H} = - {\hbar^2 \over 2\mu} {d^2\over dx^2} + D (e^{-{2x\over
d}} - 2e^{- {x\over d}} ) ~~ , \eqno(2.1) 
$$
whose eigenstates ${\cal E}$ can be associated with  $U(2)\supset
 SO(2)$ states.$^{18)}$  In order to see how this isomorphism comes
about, consider the radial equation 
$$
{1\over 2} \biggl( - {1\over r} {d\over dr} r {d\over dr} + {m^2
\over r^2} + r^2 \biggr) \phi (r) = (N+1) \phi (r) ~~ , \eqno(2.2) 
$$
which corresponds to a two-dimensional harmonic oscillator (in units where
$\hbar = \mu = e = 1)$ associated to a $U(2)$ symmetry
algebra.$^{19)}$  By carrying out the transformation 
$$
r^2 = (N+1)e^{-\rho} ~~ , 
$$
equation (2.2) transforms into 
$$
\biggl[ - {d^2 \over d\rho^2} + \biggl( {N + 1 \over 2}\biggr)^2
(e^{-2\rho} - 2e^{-\rho}) \biggr] \phi(\rho) = - m^2 \phi (\rho) ~~ ,
\eqno(2.3) 
$$
which can be identified with (2.1) after defining $x = \rho d$
and  multiplying by $\hbar^2/2\mu d^2$, provided that 
$$
\eqalignno{ 
D  & = { \hbar^2 \over 8\mu d^2} (N+1)^2 ~~ , & (2.4a)  \cr 
{\cal E} & = - {\hbar^2 \over 2\mu d^2} m^2 ~~ . & (2.4b) \cr}
$$
In the framework of the $U(2)$ algebra, the operator  $\hat N$
corresponds to the total 
number of bosons and is fixed by the potential shape according to
(2.4), while $m$, the eigenvalue of the $SO(2)$ generator $J_z$,
takes the values $m = \pm N/2$, $\pm (N-2)/2, \dots$.  The Morse
spectrum is reproduced twice and consequently for these applications
the $m$-values must be restricted to be positive.  In terms of the
  $U(2)$ algebra, it is  clear from (2.3-4) that the Morse
Hamiltonian has the algebraic realization 
$$
\hat  H = - {\hbar^2 \over 2\mu d^2} \hat J^2_z = - A \hat J^2_z ~~.
\eqno(2.5)  
$$
In addition, the $U(2)$ algebra includes the raising and lowering
operators $\hat J_+$, $\hat J_-$, which connect different energy
states in (2.3), while the angular momentum operator is given by
$\hat J^2 = {1\over 4} \hat N ( \hat N+2)$, as will be shown below.

The Morse Hamiltonian (2.5) can be rewritten in the more convenient
form  
$$
\hat H = A \hat  H^M = {A \over 2} [ ( \hat J_ + \hat J_- + \hat J_-
\hat J_+) - \hat N ] ~~ ,  \eqno(2.6)
$$
where we have used the relation $\hat J^2_z = \hat J^2 - {1\over 2} (
\hat J_+ \hat J_- + \hat J_- \hat J_+)$ and  added  the constant
term ${A \hat N^2 \over 4}$ in order to place the ground state
at zero energy.  
  The eigenfunctions of the Hamiltonian (2.7) are associated to the
$U(2) \supset  O (2)$ chain, and are given by 
$$
\matrix{ U (2) & \supset & SO (2) \cr 
\downarrow & & \downarrow \cr 
| [N] & , & \ v > \cr} 
$$
with 
$$
| [N], v> = \sqrt{ { (N - v ) ! \over N ! v ! }} \, (J_-)^v | [N], 0
> \eqno(2.7) 
$$
where $N$ is the total number of bosons fixed by the potential shape
(Eq. (2.4a)) and $v$ corresponds to the number of quanta in the
oscillator.  Both, $N$ and $v$, are related with the usual labels $j$
and $m$ of the $U(2)$ and $SO(2)$ groups, by means of$^{19)}$
$$
\eqalignno{ N & = 2 j \cr 
v & = j - m ~~ .  & (2.8)   \cr}
$$

The parameters $N$ and $A$ appearing in (2.6) are
related to the usual  
 harmonic and anharmonic constants $\omega_e$ and
$x_e\omega_e$ used in spectroscopy.$^{7)}$  
This is seen by  substituting the operator $J_z$ in (2.5) by its
eigenvalue.  
  In terms of $v$, the corresponding energy
expression  takes the form 
$$
E_M = -{A\over 2}(N+1/2) + A (N+1) (v + 1/2)  -A (v+1/2)^2 ~~ ,  \eqno(2.9)
$$
from which we immediately obtain 
$$
\eqalignno{\omega_e & = A (N+1) ~~ ,   \cr 
x_e \omega_e & = A ~~ . & (2.10) \cr}
$$
Thus, in a diatomic molecule the parameters $A$ and $N$ can be 
determined by the spectroscopic constants $\omega_e$ and
$x_e\omega_e$.

We now consider a particular molecular system.  We start by assigning
a $U^i(2)$ algebra to each relevant interatomic interaction.$^{13)}$  In
figure 1 we show the $U^i(2)$ assignment for ${\cal
D}_{3h}$-triatomic 
molecules.  All relevant operators in the model are then expressed in
terms of the generators of the molecular dynamical group, which is
given by the product 
$$
U^{1} (2) \otimes U^{2} (2) \otimes U^{3} (2) ~~ . \eqno(2.11) 
$$
A simple  realization for these generators can be given in terms
of the number  operator $\hat N_i$ and the  operators $\hat J_{\mu,
i}$ 
$$
\{ \hat N_i , \hat J_{x,i}, \hat J_{y, i} , \hat J_{z,i} \} , \quad i
= 1, 2, 3  
$$
with 
$$
\hat N_i = s^\dagger _i s_i + t^\dagger_i t_i \eqno(2.12a)  
$$
$$
\hat J_{x,i} = {1\over 2} ( t^\dagger_i s_i + s^\dagger_i t) , ~ \hat
J_{y, i} = {i\over 2} ( t^\dagger_i  s_i - s^\dagger_i t_i) , ~ \hat
J_{z, i} = {1\over 2} ( s^\dagger_i s_i - t^\dagger_i t_i ) ~~ ,
\eqno(2.12b)  
$$
where $s^\dagger_i (s_i)$ and $t^\dagger_i (t_i)$ are bosonic
operators satisfying the usual commutation relations 
$$
[s_i, s^\dagger_j ] = [ t_i , t^\dagger_j] = \delta_{ij}~~ . 
$$
All other commutators vanish.    The operators (2.12b) satisfy the
usual angular momentum commutation relations.  
Computing $\hat J^2_i = J^2_{xi} + J^2_{yi} + J^2_{zi}$, we find 
$$
\hat J^2_i = {\hat N_i \over 2} \left( {\hat N_i \over 2} + 1 \right)
~~ , 
\eqno(2.13a)
$$
from which the identification 
$$
j_i = N_i/2 ~~ , \eqno(2.13b) 
$$
is readily made.    One can  show directly from (2.12) or from
(2.13a) that $[\hat J_{\mu, i}, \hat N_i] = 0$.  The set (2.12b) thus
defines the $SU_i(2)$ subalgebra of $U_i(2)$.  Since $\hat N_i = N_i$
will remain fixed in our applications, we shall sometimes refer to
$SU_i(2)$ instead of $U_i(2)$.    
The specific boson realization (2.12) was given for reasons of
clarity, but is not necessary for the subsequent developments of the
model.  

Formally, while the vibrational 
 symmetry group of the $X_3$ molecules is ${\cal
D}_{3h}$, in practice it can be reduced to ${\cal D}_{3}$ due to the
in-plane restriction.  Since we are assigning a number to each bond
it is more convenient to work with the symmetric group $S_3$, which
is isomorphic to ${\cal D}_{3}$ through the generator identification
$$
\eqalignno{ & C_3 \leftrightarrow (123) ~~ ,  \cr 
& C^a_v \leftrightarrow (23) ~~ ,  \cr} 
$$
as indicated in Fig. 1.  The Hamiltonian of the system is then 
expanded in terms of the dynamical group generators (2.12), 
provided that we impose its invariance with respect to the symmetry
group $S_3$.   In order to explain the main features of the algebraic model
we start by considering a simple 
 form for the  Hamiltonian,  restricted to two body
interactions which preserve  the total number of quanta $V = \sum_i
v_i$, where each $v_i$ is defined as in (2.8),  
$$
\hat{\cal H} = A \sum^3_{i = 1} \hat H^M_i + {B\over 2} \sum^3_{i,j = 1
\atop i\neq j} \hat H_{ij} + {\lambda\over 2} \sum^3_{i,j = 1
\atop i\neq j} \hat V_{ij} ~~ , \eqno(2.14) 
$$
where the operators $\hat H^M_i$, $\hat H_{ij}$ and $\hat
 V_{ij}$ are  defined as 
$$
\eqalignno{ & \hat H^M_i = {1\over 2} (\hat J_{+i} \hat J_{-i} + \hat
J_{-i} \hat J_{+i}) - {\hat N_i \over 2} ~~ , &
(2.15a) \cr 
& \hat H_{ij} = 2 \hat J_{0i} \hat J_{0j} - {\hat N_i \hat N_j \over
2}  ~~ , & (2.15b) \cr 
& \hat V_{ij} = (\hat J_{+i} \hat J_{-j} + \hat
J_{-i} \hat J_{+j}) ~~ . & (2.15c) \cr} 
$$
The first term in (2.14) corresponds to three equivalent Morse
oscillators (2.6), while the two terms  $\hat
H_{ij}$ and $\hat V_{ij}$  correspond
to interactions diagonal in the chains associated to the couplings  
$$
SU^{(i)} (2)  \otimes  SU^{(j)} (2)  \supset
  SO^{(i)} (2)  \otimes  SO^{(j)} (2)   \supset SO^{(ij)}(2) ~~,
\eqno(2.16a)  
$$
$$
SU^{(i)} (2) \otimes SU^{(j)} (2)  \supset SU^{(ij)}
(2)  \supset  SO^{ij}(2) ~ ~ ,  \eqno(2.16b)
$$
respectively. The notation $SU^{(ij)}(2)$ indicates the usual
angular momentum coupling of the $SU^{(i)}(2)$ and $SU^{(j)}(2)$
states.

The basis arising from  three couplings of the form (2.16a) is
refered to as the local basis, since the Morse oscillators are
diagonal when the three $SO^i(2)$ algebras are well
defined.$^{10,19)}$   
  It should be noted that for 
most calculations,   higher order terms
are required in  the Hamiltonian (2.14) 
 in order to attain  higher  accuracy,
as we shall see in the following sections.  The physical
interpretation of these interactions will also be explained.

Once we have established the form of the Hamiltonian, we need a basis
to carry out its diagonalization.  Since the Hamiltonian is invariant
under the symmetry group $S_3$, its eigenfunctions span irreducible
representations (irreps) of this group for any given basis.  It is
convenient, however, to define a physical basis in order to classify
the states with the usual normal mode labels, as well as to simplify
the calculations.

\vfill
\eject

\noindent
{\ftit 3. Symmetry Adapted  Normal Basis} 

\

The simplest basis to diagonalize the Hamiltonian (2.14) is the one
associated to the local mode chain$^{19)}$ 
$$
\matrix{ & U^{(1)} (2) \ \otimes & U^{(2)} (2) \ \otimes & U^{(3)} (2)
\ \supset  & SO^{(1)} (2)  \otimes & SO^{(2)} (2) \ \otimes & SO^{(3)} (2)
\supset & SO(2) \cr 
& \downarrow & \downarrow & \downarrow & \downarrow & \downarrow &
\downarrow & \downarrow \cr 
| & [N_1] \ , & [N_2] \ , & [N_3] \ ; & v_1 \ , & v_2 \ , & v_3 \ ; & V >
~~ , \cr}  
$$
where below each group we have indicated the eigenvalues that label
their irreps.  Explicitly this basis is given by, 
$$
|[N_1], [N_2], [N_3]; v_1 v_2 v_3 > = | [ N_1 ]; v_1 > \, | [N_2];
v_2 > \, | [N_3 ];  v_3 > \eqno(3.1) 
$$
where $[[N_i]; v_i]$ are given by (2.7). 
 The index $v_i$ corresponds to the number of quanta in
the $i$-th oscillator, which is related to the eigenvalues $j_i$ and
$m_i$ of the $\hat J^2$ and $\hat J_{z,i}$ operators by means of 
expressions (2.8) and (2.13)
$$
v_i = j_i - m_i ~~ .  \eqno(3.2) 
$$
Since 
$$
j_i = {N_i \over 2} ~~ ,  \quad m_i \geq 0 ~~ ,  \eqno(3.3) 
$$
we find 
$$
v_i = 0, 1, 2, \dots ,  [N_i/2] ~~ . \eqno(3.4) 
$$
where $[x]$ indicates the integer part of $x$.  As mentioned above, 
 $V$ corresponds to the total number of quanta 
$$
V = \sum^3_{i =1} v_i ~~ , \eqno(3.5) 
$$
which is  conserved by the interactions in (2.14)

The contributions to the Hamiltonian (2.14) involving $SO(2)$ 
operators are diagonal in the basis (3.1)   
$$
\eqalignno{ & 
<[N_1],[N_2],[N_3]; v_1, v_2, v_{3}; V| \hat H^M_i |
[N_1],[N_2],[N_3]; 
v_1, v_2, v_{3}; V> = -v^2_i + N_i v_i ~~ , & (3.6)  \cr 
& <[N_1],[N_2],[N_3];  v_1, v_2, v_3 ; V | \hat
 H_{ij} | [N_1],[N_2],N_3] ; v_1, v_2, v_3; V>
\cr 
& \qquad \qquad =  2v_iv_j - (v_i N_j + v_j N_i) ~~ ,
& (3.7) \cr} 
$$
while the $\hat V_{ij}$ operator  has only 
non-diagonal matrix elements,  since it involves the raising
and lowering operators in the $i,j$ indices,  
$$
\eqalignno{ <[N_1], [N_2], [N_3]; & v^\prime_1, v^\prime_2,
v^\prime_3 ; V | \hat V_{ij} | [N_1], [N_2], [N_3]; v_1, v_2,
v_3 ; V> \cr 
& =  \sqrt{ v_j (v_i + 1) (N_i - v_i) (N_j - v_j +1)} \delta_
{v^\prime_i, v_i+1 ~ \delta v^\prime_j ; v_j-1} \cr 
& + \sqrt{ v_i (v_j +1) (N_j - v_j) (N_i - v_i +1) } \delta_{
v^\prime_i, v_i - 1 ~ \delta v^\prime_j, v_j +1} ~~ , & (3.8) \cr} 
$$
with $i,j = 1, 2, 3$ and $i<j$.  
Note that
because of the symmetry of the ${\cal D}_{3}$ system the same number
of bosons $N_i = N$ ({\it i.e.} the same  potential depth) is
assigned to all three bonds.  
  In the next Section the
 physical meaning of these interactions will become clear. 
  These analytical  results for the matrix
elements  and analogous ones for higher order interactions 
constitute one of the main advantages of the model.  We
 point out, however, that although the local basis is 
  convenient  from a numerical point of view, it 
 does not span  the irreps of $S_3$.  A better 
  way to carry out the diagonalization of (2.14) is to
symmetrize the local basis (3.1), for which we
  can either symmetry- 
 project the wave functions arising from the 
 local basis once the Hamiltonian has  been  diagonalized, 
or  generate the symmetry adapted  one-phonon states and then
construct 
the higher-phonon  states  by means of coupling coefficients.$^{13)}$
 For our purposes it is better to follow the second route,
since in this way the wave 
functions explicitly carry the normal labels from the outset.   In
the case where 
spurious modes are present, the building-up  procedure is essential, 
since in this way the unphysical modes can be exactly eliminated
from the space.$^{13)}$

In order to construct the  normal basis we start by establishing
explicit forms for the irreps of the group
$S_3$.  For practical reasons it is convenient to work with real
representations, so we have chosen the cartesian  harmonics as a basis
 for the  $E$ representation.  In Table I we show the character table of
the $S_3$ group, including at the right the basis functions spanning
the irreducible representations.  In Table II we indicate the
explicit irrep $E$ carried by these functions in the reference frame
of Figure 1.

We now consider the one-phonon local functions.  In this case the
basis (3.1) has the form  
$$
\eqalignno{ & | 1 > \equiv  | [N], [N], [N]; 100; 1 >  ~~ , \cr 
& | 2> \equiv | [N], [N], [N] ; 010; 1>  ~~   , \cr 
& | 3> \equiv | [N], [N], [N] ; 001; 1>  ~~ , & (3.12) \cr } 
$$
which can be readily projected to the normalized states 
$$
\eqalignno{ & ^1 \psi ^{A_1} = {1 \over \sqrt{3}}  \biggl\{ | 1> + |
2> + | 3> \biggr\} ~~ ,  \cr 
& ^ 1\psi^E_1 = {1\over \sqrt{6}} \biggl\{ 2 | 1> - | 2 > - | 3>
\biggr \} ~~ ,  \cr 
& ^1 \psi ^E_2 = {1\over \sqrt{2}} \biggl\{ | 2 > - | 3 > \biggr\} ~~
, & (3.13) \cr} 
$$
where we have used the notation $^V \Psi^\Gamma_\gamma$ for the wave
functions.  For higher phonon number the states are obtained through
the coupling$^{13)}$
$$
^{V_1 + V_2}\Psi^\Gamma_\gamma = \sum_{\gamma_1, \gamma_2} C(\Gamma_1
\Gamma_2 \Gamma; \gamma_1 \gamma_2 \gamma) ~
^{V_1}\Psi^{\Gamma_1}_{\gamma_1} ~ ^{V_2}\Psi^{\Gamma_2}_{\gamma_2}
~~ , \eqno(3.14) 
$$
where  the coupled wave functions now correspond to a
total number of phonons $V = V_1 + V_2$.  The coupling
(Clebsch-Gordan) coefficients $C(;)$ can be found in tables,$^{20)}$
or, in order to avoid phase inconsistencies, computed  in a
strightforward way using the explicit irrep given in Table II.$^{21)}$  In
Table III we present the Clebsch-Gordan coefficients derived in this
fashion. 

Using (3.14) repeatedly leads to a building-up  procedure to derive
the symmetry adapted basis for higher-phonon numbers.  To achieve
this task, 
 however, we must obtain  the decomposition of the
products $(A_1)^{v_{A_1}}  (E)^{v_{E}}$,  where $v_{A_1}$, 
$v_E$ correspond to the number of phonons in the normal modes. In
Table IV we indicate, as an example, the reductions for 
two and  three
quanta.    The procedure to obtain these reductions is a
standard one, explained in many group theory textbooks.$^{21)}$ 
 One should  bear in mind,  however, that  the boson
nature of the vibrations implies that only symmetrized products are
allowed for phonons  
 in the same mode.  For example, for two quanta in the $E$ mode
we have in general the  following reduction in terms of the symmetric
[ \ ] and antisymmetric $\{ \ \}$ contributions$^{21)}$ 
$$
E \otimes E = [E \otimes E] \oplus \{ E \otimes E\} ~~ ,
$$
where  
$$
[E \otimes E] = A_1 \oplus E  ~~ ; \quad  
 \{ E \otimes E \} = A_2  ~~ . 
$$
Since the two phonon state (3.14) associated to the product 
 $\{E \otimes E\}$ vanishes automatically, we are left only with the
symmetrized  product $[E \otimes E]$. 

The general  procedure  is now  clear.  Once the form of 
 the Hamiltonian has been  
determined by symmetry considerations, 
 we proceed to construct the symmetry adapted basis by
projecting the one-phonon local functions. 
 The higher-phonon functions are then generated from  the one-phonon
 symmetrized states   by means of 
 the coupling (3.14).
 Finally, we carry out the diagonalization in the symmetrized basis,
where  full advantage can be taken  of 
group-theoretical properties.  In particular, the Hamiltonian matrix
separates into 
blocks corresponding to the irreps of the symmetry group $S_3$.  For
example, from Table IV we  see that in the three phonon manifold
the number of functions is 10,  which  reduce to
three blocks of dimensions $3\times 3$, $1 \times 1$ and $3\times 3$,
corresponding to the irreps $A_1$, $A_2$ and $E$, respectively.
The simplification becomes  more significant as the complexity of the
molecular system and/or the phonon number increase.$^{13)}$

\

\

\noindent
{\ftit 4.  Analysis of Interactions} 

\

We now proceed to analyze  the interactions involved in
the Hamiltonian (2.14).  In order to do so it is convenient to recall
 the standard  labeling of 
states as well as the vibrational Dunham expansion 
 for  ${\cal D}_{3h}$-triatomic molecules.$^{2)}$

  As is well known, this  type of molecules exhibit three vibrational
degrees of freedom, which give rise to two 
normal modes associated to  $A_1$ and $E$ symmetries.  The normal
states  are then specified by the number of quanta in each mode
$|v_{A_1}, v_E>$.  In addition, the double degenerate  
 $E$ mode  carries an intrinsic angular momentum $l$,
whose  values depend on $v_E$ and are given by$^{2)}$ 
$$
l = v_E , ~ v_E - 2 , \dots , 1 \quad {\rm or } \quad 0 ~~ .
\eqno(4.1) 
$$
The states are then  specified  by  the
quatum numbers $v_{A}, ~ v_E$ and $l$ with the  notation 
$$
| v_{A_1} , ~  v^l_E > ~~ . \eqno(4.2) 
$$
The simplest way to reproduce the general features of the spectrum is
 by means of
 a Dunham expansion, which up to quadratic terms  takes the
form$^{2)}$  
$$
\eqalignno{ E_v (v_{A_1}, v_E, l) & = E_0 + \omega^{A_1}_{e} (v_{A_1} +
1/2) + \omega^E_e (v_E + 1) \cr 
& - x_e\omega_e^{A_1} (v_{A_1} + 1/2)^2 - x_e \omega_e^E (v_E + 1)^2
\cr 
& + x_{12} (v_A + 1/2) (v_E + 1) + g_{22} l^2 ~~ , & (4.3) \cr} 
$$
The first two 
terms in the sum  correspond to the harmonic
contributions to the energy, while  the next three terms,  $(v_A + 1/2)^2$,
$(v_E + 1)^2$ and $(v_{A_1} + 1/2)$ $(v_E + 1)$, represent the first
anharmonic corrections.  The  last term is the
intrinsic  (or vibrational) 
angular momentum  and gives rise to the correct
ordering for  states with the same value of the quantum numbers $v_A$
and $v_E$.  
 In addition, each wave function $|v_{A_1}, v^l_E > $ carries a
definite symmetry, which is closely related to the $l$ quantum
number.  For $l=0$ the states are totally symmetric and  labeled as  
$a_1$ (we use lower case letters to denote the symmetry).  For $l=
3k$, $k = 1, 2, 3, \dots ,$  two levels corresponding  to $a_1$ and
$a_2$ symmetries appear, while for $l = 3k +1$ or $l = 3k+2$, $k = 0,
1, 2, \dots$, the states exhibit  $e$ symmetry.  Note that for $l
\neq 0$ there are two components $\pm l$, although  this is not
explicit in the notation.$^{2)}$  We remark that the expansion (4.3) does
not remove the degeneracy of the levels $a_1$ and $a_2$ associated to
the $l = 3k$ states.  The same is true for any order in the Dunham
expansion.  Experimentally this degeneracy is not present, but this
cannot be taken into account by such simple parametrizations. 

Let us now analyze the interactions involved in (2.14). If we compute
the matrix elements of 
the  operators  $\sum \hat H_{ij}$ and $\sum \hat V_{ij}$ 
in the one-phonon manifold, we obtain 
$$
\eqalignno{ &< ^1 \psi^{A_1} | \sum\hat  H_{ij} |^1 \psi ^{A_1} > =
-2N  ~~ ,  &
(4.4a) \cr 
&< ^1 \psi^{A_1} | \sum\hat V_{ij} |^1 \psi ^{A_1} > = 2N ~~ ,  &
(4.4b) \cr 
& < ^1 \psi ^E_\gamma | \sum\hat  H_{ij} | ^1 \psi ^E_\gamma > = -2 N
~~ , \qquad \gamma = 1, 2 ~~ . & (4.5a) \cr 
& < ^1 \psi ^E_\gamma | \sum\hat  V_{ij} | ^1 \psi ^E_\gamma > = - N
~~ , \qquad \gamma = 1, 2 ~~ . & (4.5b) \cr } 
$$
From these results we conclude that the operator
$$
\hat {\cal H}_E \equiv - {1\over 3} \{ \sum V_{ij} + \sum H_{ij} \} 
\eqno(4.6)
$$
does not contribute to
the energy of the 
$A_1$ mode. The $-1/3$ factor was added for later convenience.   
 In Figure 2 we show the spectrum
generated by (4.6) for the two and three phonon manifolds,  as a
function of $N$.  Note that for large $N$ 
 the operator (4.6)
behaves as  $\hat n_E$, the number  of phonons in
the $E$ mode, a result we shall explain in  Section 5. 

The previous  analysis  leads to the  question of 
whether it is possible to construct from (2.15) an operator affecting
only the $A_1$ mode.  This is indeed possible and through projection
we find  
$$
\hat {\cal H}_{A_1} \equiv {1\over 3} \{ \sum \hat V_{ij} - {1\over
2} \sum \hat H_{ij} \}  ~~ , 
\eqno(4.7) 
$$
which satisfies  
$$
< ^1 \psi^E_i | \hat  {\cal H}_{A_1} | ^1 \psi^E_\gamma > = 0 ~~ ,
\quad \gamma  = 1, 2 ~~ , \eqno(4.8) 
$$
as required.  We show in Figure 3 the spectrum generated by (4.7) as
a function of $N$.  For large $N$  the 
operator  (4.7)  behaves as  $\hat n_{A_1}$, the number
 of phonons in the $A_1$ mode (see Section 5).

We have thus constructed operators that selectively affect to the
$A_1$ and $E$ modes.  The
Hamiltonian (2.4),  however, includes three independent operators.
   A  third operator can be easily derived: 
$$
\hat{\cal  V} \equiv \sum^3_{i = 1} \hat H^M_i +  { ( N -1 )
\over  2N } \sum^3_{i,j = 1 } \hat H_{ij} ~~ , \eqno(4.9) 
$$
which  is diagonal in the local basis (3.1) and satisfies 
$$
\eqalignno{ & < ^1 \psi^{A_1} | \hat {\cal V} | ^1 \psi^{A_1} > = 0
~~ , & (4.10a)  \cr 
& < ^1 \psi ^E_\gamma | \hat {\cal V} | ^1 \psi^E_\gamma > = 0 ~~ ,
\quad \gamma  = 1, 2 ~~ .  & (4.10b) \cr} 
$$
In Figure 4 we schematically 
 show the effect of ${\cal V}$  as a function of $N$ 
 in the two and three-phonon manifolds.  Note
that this  operator  is diagonal in the
local basis and  vanishes in the large $N$ limit.

 The method followed in this section 
 is general and indicates a procedure to define operators
with definite actions  over the physical space.  Additionally, the
use  of these symmetry adapted operators significantly improves the
convergence of the mean square search of parameters in the
diagonalization  of the Hamiltonian.

Although  the operators (4.6), (4.7) and (4.9) induce the
characteristics of the spectrum generated by the harmonic and
anharmonic contributions in (4.3), they cannot reproduce the effect
of the $ l^2$ term.  This  term  orders, for a positive
(negative) value of $g_{22}$, the vibrational levels in each phonon
multiplet $(v_{A_1}, v^l_E)$ according to increasing (decreasing)
value of $l$.  Figures 2-4 indicate  that the characteristic pattern of
the ${\cal D}_{3h}$ vibrational spectrum cannot be reproduced by the simple
Hamiltonian 
$$
\hat {\cal H} = \alpha \hat {\cal H}_{A_1} + \beta \hat{\cal H}_E + \gamma
\hat {\cal V} ~~. \eqno(4.11) 
$$

This analysis, however, does show that the simple algebraic
Hamiltonian (2.14) can be interpreted in a physically meaningful way
by concentrating on the symmetry properties of the interactions, as
expressed in (4.11).  
  In the next section we present a systematic procedure 
to derive the full set of interactions  in the algebraic
framework, starting from those  present in configuration space.

\  

\

\noindent
{\ftit 5.   Algebraic Interactions and  Configuration
Space Operators}

\

In order to establish the algebraic representation of
configuration-space operators, we start by analyzing the harmonic
limit of the angular momentum operators  
$$
\eqalignno{ & 
[\hat J_0, \hat J_\pm ] = \pm \hat J_\pm ~~ , & (5.1a) \cr 
& [\hat J_+, \hat J_- ] = 2\hat J_0 ~~ . & (5.1b) \cr} 
$$
The action of the $\hat J_\pm$ on the  Morse  states $|
[N], v>$ is given by 
$$
\eqalignno{ & \hat J_+ | [N], v > = \sqrt{ v (N - v + 1) } ~~ | [N], v-1 >
~~ , & (5.2a) \cr 
& \hat J_- | [N], v> = \sqrt{ ( N - v ) ( v+1) } | [N] , v + 1 > ~~ , &
(5.2b) \cr} 
$$
where $N$ and $v$ were defined in (3.2) and (3.3).  Defining the
change of scale transformation 
$$
\bar b\equiv { \hat J_+ \over \sqrt{ N}} ~~ , \qquad \bar b^\dagger \equiv
{\hat J_- 
\over \sqrt{ N} } ~~ , \eqno(5.3) 
$$
it is clear  that 
$$
\eqalignno{ & \lim_{N\to \infty} \bar b \ | [N] v> = \sqrt{ v} | [N] , v -1
> ~~ , & (5.4a) \cr 
& \lim_{N\to \infty} \bar b^\dagger | [N] v> = \sqrt{ v+1 } \ | [N] , v+1 >
~~ . & (5.4b) \cr } 
$$
which correspond to the harmonic limit of the model, as expected from
the role of $N$ in Eq. (2.4), {\it i.e.}, for infinite potential
depth the Morse potential cannot be distinguished from an harmonic
potential.    Using the definitions (3.2), (3.3) and
(5.3), we can rewrite the commutation relation (5.1a) in the new
form: 
$$
[\bar b, \bar b^\dagger] = 1 - { 2 \hat v \over N} ~~ , \eqno(5.5a) 
$$
where 
$$
\hat v = {\hat N \over 2} - \hat J _0 ~~  \eqno(5.5b) 
$$
is the Morse phonon operator corresponding to the definition (3.2).
The limit $N \to \infty$ leads to the  contraction of the $SU(2)$
algebra to the Weyl algebra generated by $b$, $b^\dagger$ and $1$,
with the usual boson commutation relation $[b, b^\dagger ] = 1$.  
Relations (5.3-5.5)  indicate the procedure to arrive at the
harmonic limit of the model.  Each $\hat J_{+i}$, $\hat J_{-i}$ should be
renormalized by dividing by $\sqrt{N_i}$ and then take the limit $N_i \to
\infty$.

As an example of this procedure we take the harmonic limit of the
 Morse Hamiltonian  (2.7) 
$$
\eqalignno{ \lim_{N\to \infty} & {1\over N} \hat H^M =
\lim_{N\to \infty} {1\over N} \left[  {1\over 2} \left( \hat J_- \hat
J_+ + \hat J_+ \hat J_- \right) - { \hat N \over 2} \right] \cr 
& = {1\over 2} ( b^\dagger b + bb^\dagger) -{1\over 2} \cr 
& =  b^\dagger b ~~ , & (5.6) \cr}
$$
which has eigenvalues $n_b$, in agreement with the harmonic limit
of Eq. (2.9).  Applying the same procedure to the symmetry projected
interactions of Eq. (4.11) we find 
$$
\eqalignno{ \lim_{N\to \infty} {1\over N} \hat{\cal H}_{a_1} & = 
\hat n_{A_1} ~, \cr 
\lim_{N\to \infty} {1\over N} \hat{\cal H}_E & =  \hat n_E ~, \cr
\lim_{N\to\infty} {1\over N} \hat{\cal V} & = 0 ~, & (5.7) \cr}
$$
where $\hat n_{A_1}$ and $\hat n_E$ are the operators corresponding
to   the number of  
phonons in the $A_1$ and the $E$ modes, respectively,  as can be
readly shown using the technique discussed below. 

We can now interpret Eq. (5.3) in the opposite sense, i.e. as a
 way  to construct the anharmonic representation of
harmonic operators.  Any given function of $b$, $b^\dagger$ can be
mapped into the same function of $\hat J_+$, $\hat J_-$ through the
correspondence 
$$
b \to {\hat J_+ \over \sqrt{N}} ~ , ~~ b^\dagger \to {\hat J_- \over
\sqrt{N}} ~~. \eqno(5.8) 
$$
As an example, we  consider again the one-dimensional harmonic
oscillator  
$$
\hat H = {1\over 2} (b^\dagger b + b b^\dagger ) ~~ , \eqno(5.9) 
$$
with eigenvalues $E = n_b+ 1/2$, and follow the reverse order in
(5.6). To  obtain its anharmonic 
representation  we  carry out the
correspondence (5.8) to get 
$$
\eqalignno{ \hat H & \to {1\over 2N} ( \hat J _- \hat J_+ + \hat J_+
\hat J_- ) \cr 
& \qquad = {1\over N} (\hat J^2 - \hat J^2_0) = \hat v + 1/2 - \hat
v^2/N ~~ , & (5.10) \cr} 
$$
which is the algebraic realization of the Morse oscillator, as shown
in  Section 2.

The general procedure to derive the algebraic realization of a given
configuration-space 
operator  is thus the following.  We  first write down 
the operator in terms of normal coordinates and momenta $\{ q,
p \}^\Gamma_\gamma$, and express it  in terms of the
harmonic bosons 
$$
\eqalignno{ 
b^{\Gamma \dagger}_\gamma & = {1\over \sqrt{2}}  (q - i
p)^\Gamma_\gamma  ~~ , \cr 
 b^{\Gamma}_\gamma & = {1\over \sqrt{2}} (q + i p  )^\Gamma_\gamma 
 ~~ , & (5.11) \cr}
$$
where $\Gamma$ denotes the irrep spanned by the normal coordinate and
$\gamma$ is its row. 
In the next step we write down the normal operators $\{
b^\dagger_{\Gamma_\gamma} ,
b_{\Gamma_\gamma} \}$ in terms of local ones  $\{ b^\dagger_i,
b_i \}$.  The 
explicit relations are of the general form 
$$
\eqalignno{  & b^{\Gamma\dagger}_\gamma = \sum_i \,
\alpha^\Gamma_{\gamma, i} b^\dagger_i \cr 
& b^\Gamma_\gamma = \sum_i \, \alpha^\Gamma_{\gamma, i} b_i 
 ~~ , & (5.12) \cr } 
$$
where the set $\{ \alpha^\Gamma_{\gamma,i} \}$ can be obtained by
projecting the 
local operators $\{ b^\dagger_i \}$ on the irrep spanned by the
normal coordinates $\{ q,p \}_\gamma^\Gamma$.  Finally, we substitute
(5.12) in 
the expression of the interaction given in terms of the normal
operators (5.11) and carry out the correspondence (5.8).  As an
example of this procedure we derive the algebraic form of the $\hat l^2$
interaction in (4.3). 

The representation of the operator $\hat l$ in terms of normal coordinates
$q_1^E$, $q_2^E$, which span the irrep $E$, is given by$^{22)}$ 
$$
\hat l = -i \left( q_1^E {\partial \over \partial q_2^E} - q_2^E {\partial
\over \partial q_1^E} \right) ~~ . \eqno(5.13) 
$$
Introducing the operators (5.11), this expression transforms into 
$$
\hat l = -i (b^{E\dagger}_{ 2} b^E_ 1 - b^{E\dagger}_{1} b^E_2 ) ~~ .
\eqno(5.14) 
$$
We now write the normal operators $\{ b^{E\dagger}_{1},
b^{E\dagger}_{2} \}$ in terms of the local ones $\{ b^\dagger_i \}$.
This can be done using the projected functions (3.13) 
$$
\eqalignno{ & b^{E\dagger}_{ 1} = {1\over \sqrt{6}} (2b^\dagger_1 -
b^\dagger_2 - b^\dagger_3 ) ~~ ,  \cr 
& b^{E\dagger}_{2} = {1\over \sqrt{2}} ( b^\dagger_2 - b^\dagger_3)
~~ , & (5.15) \cr} 
$$
and equivalent expressions for the annihilation operators.  Finally,
we substitute (5.15) into (5.14) to obtain 
$$
\hat l_{A_2} = {i\over \sqrt{3}} \{ b^\dagger_1 (b_2 - b_3) + b^\dagger_2
(b_3 - b_1) + b^\dagger_3 (b_1 - b_2) \} ~~ , \eqno(5.16) 
$$
where we have  explicitly indicated the irrep carried by the $\hat l$
operator.  The fact that it corresponds to an $A_2$ symmetry can be
deduced either by analyzing the 
transformation of $\hat l$ under the $S_3$ group or by identifying the
Clebsch-Gordan coefficients $C(EEA_2; \gamma_1\gamma_2 1)$ in (5.14).  The
corresponding realization in the  model is then obtained by
applying the correspondence (5.8)
$$
\hat l_{A_2} = {i \over N\sqrt{3}} \{ \hat J_{-1} ( \hat J_{+2} -
\hat J_{+3} ) + \hat J_{-2} 
(\hat J_{+3} - \hat J_{+1} ) + \hat J_{-3} ( \hat J_{+1} - \hat
J_{+2} ) \} ~~ .  \eqno(5.17)  
$$
In turn, the $\hat l^2_{A_2}$ operator is obtained by squaring
(5.17).  The same kind of analysis can be applied to  arbitrary 
configuration space interactions.$^{22)}$ 

We have  presented in this section  a general method to derive the 
realization of  operators in the algebraic  model, starting from
their 
representation in configuration space.  This  procedure  considerably
increases the power of the algebraic approach, since it can be used
to  incorporate into the model the fundamental interactions known
from the  configuration space  methods. 
 Note that this procedure allows,
in principle, to establish the relation  between the algebraic
parameters and the force field strengths obtained from {\it ab
initio} calculations.   It is  also
possible, however,  to  apply the 
model in a purely algebraic fashion and still deduce the fundamental
interactions, as we explain in the next Section. 

\

\

\noindent
{\ftit 6.   The Anharmonic Oscillator Symmetry Model} 

\

In this Section we present a general framework  to construct all
interactions in the algebraic model in a systematic way.  We shall
henceforth refer to this  procedure, together with the methods
introduced in the previous sections, as the Anharmonic Oscillator
Symmetry Model (AOSM).

We start by 
introducing  a set of generators with well-defined tensorial properties
under the point group 
$$
\hat J^\Gamma_{\mu,\gamma} = \sum_i \alpha^\Gamma_{\gamma, i} \hat
J_{\mu, i} ~~ , \eqno(6.1) 
$$
where $\mu = +, -, 0$.  For the case of ${\cal D}_{3h}$ molecules the
 expansion coefficients are the same as those in Eq. (5.12).
  We then  construct from these symmetry projected generators a set
of interactions that are scalars under the point group, such as 
$$
\eqalignno{ &  T^{A_1}_{\pm} (\Gamma) = {1\over 2}  \sum_\gamma
\left( \hat 
J^\Gamma_{-,\gamma} 
\hat J^\Gamma_{+, \gamma} + \hat J^\Gamma_{+,\gamma} \hat
J^\Gamma_{-, \gamma} \right) ~~ , & (6.2a) \cr 
\noalign{\hbox{and}} 
&  T^{A_1}_0 = (\Gamma)  \sum_\gamma \hat J^\Gamma_{0,\gamma} \hat
J^\Gamma_{0,\gamma} ~~ .  & 
(6.2b) \cr} 
$$
Higher order tensors can be systematically constructed by means of
(6.1) and the Clebsch-Gordan coefficients for the point
group. 

For triatomic ${\cal D}_{3h}$-molecules $\Gamma = A_1, E$,  the
relevant symmetry projected generators are 
$$
\eqalignno{ & \hat J^{A_1}_{\mu, 1} = {1\over \sqrt{3}} \left( \hat
J_{\mu, 1} + \hat J_{\mu, 2} + \hat J_{\mu, 3} \right) ~~ ,  \cr 
& \hat J^E_{\mu, 1} = {1\over \sqrt{6}} \left( 2 \hat J_{\mu, 1} -
\hat J_{\mu, 2}  - \hat J_{\mu, 3} \right) ~~ ,  \cr
& \hat J^E_{\mu, 2} = {1\over \sqrt{2}} \left( \hat J_{\mu, 2} - \hat
J_{\mu, 3} \right) ~~ , & (6.3) \cr}
$$
with $\mu = +, -, 0$.  According to (6.2) we can construct four
possible interactions that are quadratic in $\hat J^\Gamma_{\mu,
\gamma}$, three of which correspond to linear combinations of the
terms in (2.14)
$$
\eqalignno{  T^{A_1}_\pm (A_1) & = {3\over 2} N + (\sum^3_{i =1}
H^M_i + {1\over 2} \sum^3_{i,j=1\atop i\neq j} \hat V_{ij} ) & (6.4a) \cr
 T^{A_1}_\pm (E) & = 2 N + {1\over 3} (4\sum^3_{i =1}
H^M_i - \sum^3_{i,j=1\atop i\neq j} \hat V_{ij} ) & (6.4b) \cr
 T^{A_1}_0 (A_1) & = {9\over 4} N^2 - (\sum^3_{i =1}
H^M_i - \sum^3_{i,j=1 \atop i\neq j} \hat H_{ij} ) ~~ ,  & (6.4c) \cr}
$$
while the fourth is not independent, since 
$$
\vec J^2_{A_1} + \vec J^2_E = \sum^3_{i = 1} \vec J^2_i = {3\over 4}
 \hat N (\hat N + 2) ~~ . \eqno(6.4d) 
$$
In addition to the operators in (6.2) which  transform as $A_1$ under
${\cal D}_3$, we can also construct other bilinear combinations with
well-defined tensor properties, 
$$
\eqalignno{ & \hat T^E_1 = {1\over 2} \left( \hat J^E_{-, 2} \hat
J^E_{+, 2} - \hat J^E_{-, 1} \hat J^E_{+, 1} \right) ~~ , \cr  
& \hat T^E_2 = {1\over 2} \left( \hat J^E_{-, 1} \hat J^E_{+, 2} +
\hat J^E_{-, 2} \hat J^E_{+, 1} \right) ~~ , \cr 
& \hat T^{A_2} = {i \over 2} \left( \hat J^E_{-, 1} \hat J^E_{+, 2}
- \hat J^E_{-, 2} \hat J^E_{+, 1} \right) ~~ . & (6.5) \cr} 
$$
The  operator $\hat T^{A_2}$ 
 is proportional to the intrinsic angular momentum
operator of (5.17). 

In lowest order ({\it i.e.} quadratic in $\hat T^\Gamma_\gamma$)
there are two possible $A_1$ interactions, $(\hat T^E_1)^2 + (\hat
T^E_2)^2$ and $(\hat T^{A_2})^2$.  In order to interpret these
interactions we take the harmonic limit of (6.5) 
$$
\eqalignno{ & \lim_{N\to \infty} {1\over N} \hat T^E_1 = \hat L_x =
{1\over 2} \left( b^\dagger_{E_2} b_{E_2} - b^\dagger_{E_1}
b_{E_1}\right)~~  , \cr 
& \lim_{N \to \infty} {1\over N} \hat T^E_2 = \hat L_y = {1\over 2}
\left(b^\dagger_{E_1} b_{E_2} + b^\dagger_{E_2} b_{E_1} \right) ~~ , \cr 
& \lim_{N\to \infty} {1\over N} \hat T^{A_2} = \hat L_z = {i \over 2}
\left( b^\dagger_{E_1} b_{E_2} - b^\dagger_{E_2} b_{E_1} \right) ~~ .
& (6.6) \cr} 
$$
The operators $\hat L_x$, $\hat L_y$ and $\hat L_z$ close under the
commutation relations of $SU (2)$.  In the harmonic limit $(\hat
T^{A_2})^2$ corresponds to $\hat L^2_z = \hat l^2 /4$ (see eq.
(5.14)) while $(\hat T^E_1)^2  + (\hat T^E_2)^2$ goes to  $\hat L^2_x
+ \hat L^2_y = \vec L^2 - \hat L^2_z$ with $\vec L^2 = \hat n_E (\hat
n_E + 2)/4$, corresponding to an anharmonic contribution to the $E$
mode.  

In the next order ({\it i.e.} cubic in $\hat T^\Gamma_\gamma$) we
can first couple $T^E_\gamma$ to $E$ 
and then couple again  the resulting operator to $T^E_\gamma$
 to obtain an $A_1$ operator, 
$$
\eqalignno{ 
\hat {\cal O}^{A_1} & = \left( \hat T^E_2 \hat T^E_2 - \hat T^E_1
\hat T^E_1 \right) \hat T^E_1 + \left( \hat T^E_1 \hat T^E_2 + \hat
T^E_2 \hat T^E_1 \right) \hat T^E_2  \cr
& = - {1\over 2} \left( \hat T^3_+ + \hat T^3_- \right) ~~ ,   &
(6.7) \cr} 
$$
where we have introduced $\hat T_{\pm} = \hat T^E_1 \pm i \hat
T^E_2$, which in the harmonic limit reduce to the ladder operators
$\hat L_\pm = \hat L_x \pm i \hat L_y$.  In the harmonic limit 
the operator (6.8) has a clear physical interpretation:  it couples
states with $\Delta L_z = \pm 3$, or expressed in terms of the
intrinsic angular momentum, it couples states with the same $v_E$ and
$\Delta l = \pm 6$.  For this reason it 
splits the $a_1$ and $a_2$ vibrations that are associated with the
$(v_{A_1}, v^{l = 3}_E)$ multiplet. 

The explicit realization of $\hat {\cal O}^{A_1}$ in the AOSM can be
obtained by expressing $\hat T_+$ (and $\hat T_- = (\hat T_+)^\dagger$)
in terms of the $\hat J_{\pm, i}$ operators through equations (6.5)
and (6.3).  The final result is 
$$
\eqalignno{ \hat T_+ & = - {1\over 6} \biggl[ 2 \hat J_{-1} \hat
J_{+1} - \hat J_{-2} \hat J_{+2} - \hat J_{-3} \hat J_{+3} \cr 
& \qquad - \hat J_{-1} ( \hat J_{+2} + \hat J_{+3} ) - \hat J_{-2} (
\hat J_{+1} - 2 \hat J_{+3} \cr 
& \qquad - \hat J_{-3} ( \hat J_{+1} - 2 \hat J_{+2} ) \biggr] \cr 
& + {i \over 2 \sqrt{3}} \biggl[ \hat J_{-1} ( \hat J_{+2} - \hat
J_{+3} ) + \hat J_{-2} ( \hat J_{+1} - \hat J_{+2} ) \cr 
& \qquad + \hat J_{-3} ( \hat J_{+3} - \hat J_{+1} ) \biggr] ~~ . & (6.8)
\cr} 
$$
In the next section we show that the interaction (6.7) is essential
to describe the highly anharmonic molecule $H^+_3$.  It should be
clear that the AOSM can be applied in a similar way to molecules
exhibiting arbitrary symmetry groups.  We
 have thus presented a  systematic
 procedure to construct, up to a certain order, all relevant
interactions, based on the introduction of operators
with well-defined tensorial properties under the point group (see
e.g. equation (6.1)), which can then be combined into ${\cal D}_{3h}$
scalar interactions.

\

\

\noindent
{\ftit 7.  Application to H$^+_3$, Be$_3$ and Na$^+_3$ }

\

In this section we apply the  AOSM  to Na$^+_3$, Be$_3$
and H$^+_3$.  As mentioned in the Introduction, we have chosen these
molecules because they 
exhibit a wide range of behavior, ranging from the very anharmonic
spectrum of H$^+_3$ to the almost exact harmonic spectrum of
Na$^+_3$. 

According to the discussion presented in the previous section, 
 the Hamiltonian 
$$
\hat {\cal H} = \alpha \hat {\cal H}_{A_1} + \beta \hat {\cal H}_E +
\gamma \hat {\cal V} + \delta \hat l^2 ~~ ,  \eqno(7.1) 
$$
contains the main physical interactions that describe a ${\cal
D}_{3h}$-triatomic molecule, whose spectrum is close to the one
generated by the Dunham expansion 
 (4.3).  As mentioned before,  the Dunham   expansion
   implies  a degeneracy between the $a_1$ and $a_2$ levels
associated to the same quantum number $l$, while the spectrum
generated by (7.1) does lead to their splitting, although it is
generally small.  Experimentally this 
splitting is observed, even for molecules like Na$^+_3$, in spite of
 its  harmonic behavior.  In the model Hamiltonian (7.1)
this effect is 
produced by the $\hat {\cal H}_{A_1}$ operator, as can be seen from Fig.3 .

We now consider the molecules H$^+_3$, Be$_3$ and
Na$^+_3$.  The atoms in the first molecule are very light and the
spectrum is highly anharmonic,  a
fact that is reflected by a strong splitting of the $a_1$, $a_2$
levels  (200 
cm$^{-1}$ for $V=3$), as well as by a relatively large splitting of
levels  belonging to the same $(v_{A_1}, v_E)$ multiplet but with
different value of $l$  (220 cm$^{-1}$ for $V=2$).    This is in
contrast with the case of 
Na$^+_3$, where the splitting between the $a_1$ and  $a_2$ levels
is very small (0.11 cm$^{-1}$ for $V=3$) as well as the $l$-dependent
splitting (0.82 cm$^{-1}$ for $V=2$).  
 The molecule Be$_3$ exhibits
an intermediate behavior, although in this case an $a_1$, $a_2$
splitting is not present in the fitted data, since we have generated
its spectrum from an {\it ab initio} calculation where no splitting
terms are included.$^{15)}$

In Table V we present a least square fit calculation for
H$^+_3$, Be$_3$ and Na$^+_3$ up to three quanta, using the
Hamiltonian (7.1) with $\delta = 0$.  The standard deviation (rms) was
taken to be 
$$
rms = \sqrt{{\sum^n_i (E^i_{exp} - E^i_{th})^2 \over n-n_p}} ~~ ,
\eqno(7.2) 
$$
where $n$ and $n_p$ correspond to the number of fitted levels and
parameters involved, respectively.    From this
calculation we find a large  difference in the quality of the fit  between
H$^+_3$ and the other two molecules.   In Table VI we present
the same calculations, but including the $\hat l^2$ interaction.  We
see that the difference in quality  persists; while the
Hamiltonian (7.1) is quite sufficient to describe the Be$_3$ and Na$^+_3$
molecules, we clearly require additional interactions to properly
describe H$^+_3$.  This fact is in accordance with the work of
Carter and Meyer,$^{16)}$ who are forced to include twice as many
terms in the potential for H$^+_3$ than for the Na$^+_3$
molecule.   The simplest possible set of such interactions in the
AOSM  correspond to higher powers  
 of the symmetry adapted operators  (4.6) and (4.7).  We
propose 
$$
\hat{\cal H} ^{2}_{A_1} , \,\,\,   \hat{\cal H} ^{2}_{E}, \,  \,\, 
 \hat{\cal H}_{A_1E}\, \equiv \, { (\hat{\cal H}_{A_1}\hat{\cal H}_E
+ \hat{\cal H}_{E} \hat {\cal H}_{A_1}) \over 2} 
 ~~ . \eqno(7.3) 
$$
If we add this set of interactions to the  Hamiltonian (7.1) in 
the energy fit for H$^+_3$ the rms deviation  reduces  to 15.74 
cm$^{-1}$.    It is
possible to further  improve the fit by taking into account the
interaction (6.7) in addition to the set (7.3).  A more general
algebraic 
Hamiltonian  to describe ${\cal D}_{3h}$ molecules is then 
$$
\eqalignno{ \hat {\cal H} & = \alpha \hat {\cal H}_{A_1} + \beta \hat{\cal
H}_E + \gamma \hat {\cal V} + \delta \hat l^2 + \alpha^{[2]} \hat
{\cal H}^{2}_{A_1} \cr 
& \qquad + \beta^{[2]} \hat{\cal H}^2_E + \xi^{[2]} \hat {\cal
H}_{A_1E} + \epsilon (\hat T^3_+ + T^3_- ) ~~ . & (7.4) \cr} 
$$
As mentioned before, the 
 operator $\hat T^3_+ + \hat T ^3_-$ has the effect of splitting the $a_1$
and 
$a_2$ levels arising  from the same angular momentum $l$, which 
explains the need for this interaction in H$^+_3$.  In Table
VII we present 
the least square energy fit to  H$^+_3$   using the
Hamiltonian (7.4), with an rms deviation of 5.84 cm$^{-1}$.   We
remark that in order to describe H$^+_3$ for higher phonon numbers we
need to include higher order interactions.   We believe, however,
that this result is very encouraging.  If we omit the purely
anharmonic interaction $\hat{\cal V}$  the rms increases to  24.37 
cm$^{-1}$, while carrying out the harmonic limit $(N\to \infty)$,
where $\hat{\cal V}=0$, the rms obtained is 31.17 cm$^{-1}$.

\

\

\noindent
{\ftit 8.  Conclusions} 

\

We have introduced the Anharmonic Oscillator Symmetry Model and
applied it  to a set of 
${\cal D}_{3h}$-triatomic molecules.  The model is based on symmetry
methods which systematically incorporate group theoretical
techniques, providing a clear methodological procedure that can be
applied to more complex molecules.  We have introduced symmetry adapted
operators that have a
specific action 
over the function space.  This is a general procedure which gives
rise to a clear physical interpretation of the interactions and has
the additional advantage of 
considerably improving the convergence during the least square energy
fit.  Furthermore, based on the harmonic limit of the SU(2) algebra
we have proposed a systematic approach to derive an algebraic
realization of interactions given in configuration space.  The model 
 surmounts one of the main objections raised against the use of
algebraic
models, where it was not possible to obtain a direct correspondence
with the configuration-space 
  approaches.  Although we have illustrated this  procedure
by means of the $\hat l^2$ interaction,  the method can
be used for arbitrary operators. 
  For the general case  when there is no
information about the 
 form  of these interactions in configuration space, we have
devised an  algebraic procedure to derive them  using their 
tensorial  structure  under the point group.  The
combination of the different methodologies leads to the AOSM, 
 which can be  applied in the same fashion to more complex molecules.
  We remark  that the model can be extended in several ways.  For
example, Fermi resonances can be taking into account using
perturbation theory, while  
 the 
rotational degrees of freedom can be incorporated by coupling the
vibrational wave 
functions to rotational states carrying the appropriate point
symmetries.$^{23,24)}$ 

We believe that the  AOSM  represents a systematic,
simple but accurate alternative to  configuration space  methods,
particularly for polyatomic molecules, where the integro-differential
approaches are too complex to be applied or require very large 
numerical calculations.  Since the model provides manageable wave
functions, it is possible to evaluate the matrix elements of
arbitrary physical operators, which have a simple representation in
the algebraic space.  A finer test for the model is to  use these
wave functions, for example, for  the evaluation of infrared
and Raman intensities.  The transition  operators can be 
constructed by applying our method to the  configuration-space 
parametrizations, which correspond to the harmonic limit, $N\to
\infty$, or purely algebraically by using their tensorial properties
under the corresponding point group.    The analysis of
electromagnetic   intensities, as well as the
application of the model to other molecular  systems will be
presented in future publications.$^{17,23)}$

\

\

\noindent
{\ftit Acknowledgments}

\

We are grateful to P. Van Isacker and F.  Iachello for his
continuous interest and 
useful comments.  This work was supported in part by 
 the  European 
Community under contract  nr CI1$^\ast$-CT94-0072, 
DGAPA-UNAM under
project IN105194,    CONACyT-M\'exico  under project 400340-5-3401E
and Spanish DGCYT under project PB92-0663.

\

\noindent
{\ftit References}

\item{ 1. }  J. Michael Hollas, {\it Modern Spectroscopy}, John Wiley
1992; Kazuo Nakamoto, Infrared and Raman Spectra of Inorganic and
Coordination Compounds.  Wiley - Interscience publication, 1978. 

\item{ 2. }  Gerard Herzberg, {\it Infrared and Raman Spectra of
Polyatomic Molecules}, van Nostrand, New York, 1950.

\item{ 3. }  E.B. Wilson, Jr., J.C. Decious and P. Cross, Molecular
Vibrations,  Dover, New York,  1980.

\item{ 4. } W.T. Raynes, P. Lazzeretti, R. Zanesi, A.J. Sadly and
P.W. Fowler, Mol. Phys. {\bf 60} (1987)  509; G.D. Carney and R.N.
Porter, J. Chem. Phys. {\bf 65} (1976) 3547.  

\item{ 5. }  D.L. Gray and A.G. Robiette, Mol. Phys. {\bf 37} 
(1979)  1901. 

\item{ 6. }  J.F. Ogilvie, J. Mol. Spect., {\bf 69} (1978)  169; W.
Meyer and  P. Botschwina, J. Chem. Phys., {\bf 84} (1986) 891.

\item{ 7. }  F. Iachello, Chem. Phys. Lett. {\bf 78} (1981)  581; F.
Iachello and R.D. Levine, J. Chem. Phys. {\bf 77} (1982)  3046.

\item{ 8. } F. Iachello, S. Oss and R. Lemus, J. Mol. Spect. {\bf
146} (1991)  56; Ibidem {\bf 149} (1991)  132.

\item{ 9. }  F. Iachello and S. Oss, J. Mol. Spect. {\bf 142}
(1990)  85. 

\item{ 10. }  O.S. van Roosmalen, I. Benjamin and R.D. Levine, J.
Chem. Phys. {\bf 81} (1984)  5986.

\item{ 11. }  F. Iachello and S. Oss, Phys. Rev. Lett. {\bf 66}
(1991)  2976;  Chem. Phys. Lett. {\bf 187} (1991)  500; 
F. Iachello and S. Oss. Chem. Phys. Lett. {\bf 205}
 (1993) 285;  J. Chem. Phys. {\bf 99} (1993)  7337.

\item{ 12. } J.M. Arias, A. Frank, R. Lemus and F. P\'erez-Bernal,
 Rev. Mex. F\ii s. {\bf 41} (1995) 728.

\item{ 13. }  R.  Lemus and A. Frank,  J. Chem. Phys. {\bf 101}
(1994) 8321; A. Frank and R. Lemus, Phys. Rev. Lett. {\bf 68} (1992)
413. 

\item{ 14. }  R. Bijker, A.E.L. Dieperink and A. Leviatan, 
 Phys. Rev. A.  {\bf 52} (1995) 2786.

\item{ 15. }  A.P. Rendell, T.J. Lee and P.R. Taylor, J. Chem. Phys.
{\bf 92} (1990)  7050.

\item{ 16. }  S. Carter and W. Meyer, J. Chem. Phys. {\bf 93} (1990)
8902.

\item{ 17. }   F. P\'erez-Bernal, R. Bijker, A. Frank, R. Lemus and
J.M. Arias, preprint, Chem-ph/96 03003.

\item{ 18. } Y. Alhassid, F. G\"ursey and F. Iachello, Ann. of Phys.
{\bf 148} (1983) 346.

\item{ 19. }  A. Frank and P. Van Isacker, {\it Algebraic Methods in
Molecular and Nuclear Structure Physics}, Wiley, New York, 1994.

\item{ 20. }  S.L. Altmann and P. Herzig, {\it Point Group Theory
Tables}, Clarendon Press, Oxford, 1994.

\item{ 21. }  M. Hamermesh, {\it Group Theory and its Applications to
Physical Problems}, Addison-Wesley, 1962.

\item{ 22. }  K.T. Hecht, J. Mol. Spect. {\bf 5} (1960) 355.

\item{ 23. }  J.K.G. Watson, J. Mol. Spect. {\bf 103} (1984) 350;
Ibid, Can. J. Phys. {\bf 72} (1994) 238; J. Tennyson and J.R.
Henderson, J. Chem. Phys. {\bf 91} (1989) 3815. 

\item{ 24. }  A. Frank, R. Lemus, J. P\'erez-Bernal and J.M. Arias,
to be published.

\vfill
\eject

\centerline{\ftit  Figure Captions}

\

\item{Figure 1.}  Assignment of the $U^i(2)$ algebras for ${\cal
D}_{3h}$ triatomic molecules and the selection of the Cartesian
coordinate system.  The elements of the symmetry group ${\cal
D}_{3}$ are also indicated.

\

\item{Figure 2.} Eigenvalues of the operator $\hat{\cal H}_E$ as a
function of the number of bosons $N$ for (a) two phonons and (b)
three phonons.

\

\item{Figure 3.}   Eigenvalues of the operator $\hat {\cal H}_{A_1}$
as a function of the number of bosons $N$ for (a) two phonons and (b)
three phonons.

\

\item{Figure 4.}  Eigenvalues of the operator $\hat{\cal V}$ as a
function of the number of bosons $N$ for two and three phonons.

\
\vfill
\eject

\

\noindent
{\bf Table I. } {Character table for  the ${\cal D}_{3}
\approx S_3$ group.  The set $\{  R_x, R_y, R_z\}$ represents the
components of an axial vector.}
$$
\matrix{ {\cal D}_{3} & E & C_3 (2) & C_2 (3) & \hbox{Basis
functions}  \cr 
\noalign{\hrule} 
A_1 & 1 & 1 & 1 &  2z^2 - x^2 - y^2 \cr 
A_2 & 1 & 1 & -1 & z; R_z ~~ , \cr 
E & 2 & -1 & 0 & (x, y) ~ , ~~ (R_x,  R_y) ; \cr 
& & & & (zx, yz) ~~ (xy, x^2 - y^2) \cr} 
$$

\

\

\noindent
{\bf Table II. } {The  $E$ irreducible
representation of the generators of the $S_3$ group in the
Cartesian basis of Table I.}
$$
\matrix{ Irrep & (123) & (23) & S_3 \cr 
& C_3 & C^a_2 & {\cal D}_{3} \cr 
\noalign{\hrule} 
E & \pmatrix{ - {1\over 2} & - {\sqrt{3} \over 2} \cr {\sqrt{3}\over
2} & - {1\over 2} \cr} & \pmatrix{ 1 & 0 \cr 0 & -1 \cr} \cr} 
$$

\

\

\noindent
{\bf Table III.} {Clebsch-Gordan coefficients
$C(\mu\nu\Gamma; ij\gamma)$ for the $S_3$ group  compatible
with the irreps of Table II.}
$$
\matrix{\mu\times \nu \hfill  & (ij)\hfill \cr
\noalign{\vskip .2truecm}
\noalign{\hrule}
\noalign{\vskip .2truecm} 
(\Gamma)_\gamma \hfill & C(\mu\nu\Gamma; ij\gamma) \hfill \cr}
\qquad 
\matrix{A_2 \times E & (11) & (12) \cr
\noalign{\vskip .2truecm}
\noalign{\hrule}
\noalign{\vskip .2truecm} 
(E)_1 & 0 & 1 \cr (E)_2 & -1 & 0 \cr}
$$

\

$$ 
\matrix{ E\times E & (11) & (12) & (21) & (22) \cr 
\noalign{\vskip .2truecm}
\noalign{\hrule}
\noalign{\vskip .2truecm} 
A_1 & 1/\sqrt{2} & 0 & 0 & 1/\sqrt{2} \cr 
A_2 & 0 & 1/\sqrt{2} & - 1/\sqrt{2} & 0 \cr
(E)_1 &  1/\sqrt{2} & 0 & 0 & -1/\sqrt{2} \cr 
(E)_2 & 0 & -1/\sqrt{2} & -1/\sqrt{2} & 0 \cr}
$$

\vfill
\eject

\centerline{
{\bf Table IV. } {${\cal D}_{3}$ decompositions  for two
and three quanta.} }

$$
\matrix{ A_1 & E & & \hbox{Dimension of the} \cr
v_{A_1} & v_E & \hbox{Irreps} & \hbox{reducible
representation}\cr 
& \cr 
\noalign{\hrule} 
& \cr 
2 & 0 & A_1 \hfill & 1 \cr 
0 & 2 & A_1 \oplus E \hfill & 3 \cr 
1 & 1 & E \hfill & 2 \cr 
& & & 6 \cr 
& \cr 
& \cr 
3 & 0 & A_1 \hfill & 1 \cr 
0 & 3 & A_1 \oplus A_2 \oplus E \hfill & 4 \cr 
2 & 1 & E \hfill & 2 \cr 
1 & 2 & A_1 \oplus E \hfill & 3 \cr 
& & & 10 \cr} 
$$

\vfill
\eject

\noindent
{\bf Table V. }  Least square energy fit for H$^+_3$, Be$_3$
and Na$^+_3$.  All energies are in cm$^{-1}$.  We 
indicate the rms deviation (7.2) and the 
 parameters obtained. The number
of bosons $N$ was taken to be 30.  The energy difference is given by
$\Delta E = E_{th} - E_{exp}$, where $E_{exp}$ are taken from {\it ab
initio}  calculations. 

$$
\matrix{ \noalign{\hbox{ \ \ \ \ \ \ \ \ \ \  \ \ \ \ \ \ \ \ \ \ \ \
 \ \ \ \ \ \ \ \ \ \ \ \ \ \  \ \ \ \  $ 
H^+_3  \ \ \ \ \ \ \ \ \ \ \ \ \ \ \ \  \ \ \ Be_3  \ \ \ \ \ \ \ \ \ \ \ \
 \ \ \ \ \ 
Na^+_3$}} \cr  
(v_{A_1} v^l_E) & \hbox{Symmetry} & Exp.^{16)} &  \Delta E &
Exp.^{15)}  & 
\Delta E & Exp.^{16)}  &  \Delta E \cr 
& \cr 
(01^1) & e & \hfill 2521.27  & \hfill -0.17 & \hfill 399.10  & \hfill
4.09 & \hfill 99.95 & \hfill 1.68 \cr 
(10^0) & a_1 & \hfill 3178.32 & \hfill -29.32 & \hfill 458.40 &
\hfill 0.25 &  \hfill 140.45 & \hfill 2.16 \cr 
& \cr 
(02^0) & a_1 & \hfill 4777.02 & \hfill 102.26 & \hfill 782.40 &
\hfill 8.33 & \hfill 198.90 & \hfill 2.02  \cr 
(02^2) & e & \hfill 4997.41 & \hfill -45.57 & \hfill 794.40 & \hfill
1.31 & \hfill 199.72 & \hfill 1.25 \cr 
(11^1) & e & \hfill 5553.67 & \hfill -14.97 & \hfill 845.10 & \hfill
2.02 & \hfill 239.29 & \hfill 1.70 \cr
(20^0) & a_1 &  \hfill 6261.92 & \hfill -22.77 & \hfill 907.60 &
\hfill -0.04 & \hfill 280.35 & \hfill 1.66 \cr 
& \cr 
(03^1) & e & \hfill 7003.49 & \hfill 105.59 & \hfill 1161.90 & \hfill
 3.91 & \hfill 297.67 & \hfill 0.25 \cr 
(03^3) & a_1 & \hfill 7282.52 & \hfill -102.30 & \hfill 1185.90 &
\hfill -13.85 & \hfill 299.26 & \hfill -1.25 \cr 
(03^3) & a_2 & \hfill 7492.64 & \hfill -105.84 & \hfill 1185.90 &
\hfill -3.95 & \hfill 297.67 & \hfill -1.36 \cr 
(12^0) & a_1 & \hfill 7769.09 & \hfill 116.14 & \hfill 1216.00 &
\hfill 9.71 & \hfill 337.19 & \hfill -0.15 \cr 
(12^2) & e & \hfill 7868.64 & \hfill -32.58 & \hfill 1228.00 & \hfill
-3.37 & \hfill 337.94 & \hfill -0.86 \cr 
(21^1) & e & \hfill 8486.90 & \hfill 1.37 & \hfill 1281.90 & \hfill
-0.72 & \hfill 378.06 & \hfill -0.91 \cr 
(30^0) & a_1 & \hfill 9251.42 & \hfill 14.89 & \hfill 1347.60 &
\hfill -1.04 & \hfill 419.70 & \hfill -1.49 \cr} 
$$

$$
\matrix{ \hfill rms   &  \ \ \ &  \ \ \ \ \ \ \ \ \  78.55   & &
& 
\ \ \ \ \ 6.48  &
& &  \ \ \ \ \ 2.58 \cr 
& \cr 
 & \ \ \ \alpha   & \hfill \ \ \ \ \ \ \ \ \  3148.996  & & &
\hfill \ \ \ 
 \ \ 458.653 & & &
\hfill \ \ \ \ \  142.608 
\cr 
\hbox{Parameters}  &   \ \ \  \beta  &  \hfill \ \  \ \ \ \ \ \
\ 2521.105   & & & 
\hfill  \ \ \ \ \ 403.185 & & & 
\hfill \ \ \ \ \ 101.633 \cr 
&   \ \ \ \gamma   & \hfill \ \ \ \ \ \ \ \ \  3796.387 & & &
\hfill  \ \ \ \ \ 328.843  & & &
\hfill \ \ \ \ \ 49.434 \cr} 
$$

\vfill
\eject

\noindent
{\bf Table VI.}  Least square energy fit for H$^+_3$, Be$_3$
and Na$^+_3$  using the  Hamiltonian (7.1).  We show
 the energy differences $\Delta E = E_{th} - E_{exp}$.  The values of
the energies   $E_{exp}$
are given in Table V. 

$$
\matrix{ & & H^+_3 & Be_3 & Na^+_3 \cr 
(v_{A_1} v^l_E) & Symmetry & \Delta E & \Delta E & \Delta E \cr 
& \cr 
(01^1) & e & \hfill -37.18 & \hfill 0.51 & \hfill 0.93 \cr 
(10^0) & a_1 & \hfill -21.70 & \hfill 0.02 & \hfill 1.95 \cr 
& \cr 
(02^0) & a_1 & \hfill -16.32 & \hfill -0.74 & \hfill 0.37 \cr 
(02^2) & e & \hfill -33.84 & \hfill 0.17 & \hfill 0.84 \cr 
(11^1) & e & \hfill -35.74 & \hfill 0.82 & \hfill 1.68 \cr 
(20^0) & a_1 & \hfill - 13.42 & \hfill -0.04 & \hfill 1.26 \cr 
& \cr 
(03^1) & e & \hfill 18.66 & \hfill - 2.05 & \hfill -1.19 \cr 
(03^3) & a_1 & \hfill 16.05 & \hfill -1.23 & \hfill -0.34 \cr 
(03^3) & a_2 & \hfill -0.62 & \hfill 0.61 & \hfill -0.33 \cr 
(12^0) & a_1 & \hfill 46.62 & \hfill 1.90 & \hfill -0.01 \cr 
(12^2) & e & \hfill -12.83 & \hfill -1.36 & \hfill 0.34 \cr 
(21^1) & e & \hfill -3.38 & \hfill 0.79 & \hfill -0.19 \cr 
(30^0) & a_1 & \hfill 22.56 & \hfill -1.66 & \hfill -2.06 \cr 
& \cr 
\noalign{\hrule} 
& \cr
& rms & 30.15 & 1.24 &  \hfill 1.25 \cr 
& \cr 
\noalign{\hrule}
& \cr 
& \alpha & \hfill 3156.616 & \hfill 458.911 & \hfill 142.396 \cr 
Parameters & \beta & \hfill 2446.638 & \hfill 396.265 & \hfill
100.317  \cr
& \gamma & \hfill 3131.825 & \hfill 209.744 & \hfill 21.312 \cr 
& \delta & \hfill -12.485 & \hfill -0.9533 & \hfill -0.1867 \cr 
& \cr
\noalign{\hrule} }
$$

\vfill
\eject

\noindent
{\bf Table VII.}  Least square energy fit of the  H$^+_3$
molecule using the Hamiltonian (7.6).  
$$
\matrix{ &  H^+_3 \cr 
& \cr 
(v_{A_1} \, v^l_E) & Symmetry & \Delta E \cr 
& \cr 
(01^1) & e & \hfill -1.55 \cr 
(10^0) & a_1 & \hfill 0.42 \cr 
& \cr 
(02^0) & a_1 & \hfill 7.48 \cr 
(02^2) & e & \hfill -5.69 \cr 
(11^1) & e & \hfill -0.61 \cr 
(20^0) & a_1 & \hfill -0.11 \cr 
& \cr 
(03^1) & e & \hfill -4.46 \cr 
(03^3) & a_1 & \hfill 3.18 \cr 
(03^3) & a_2 & \hfill 2.44 \cr 
(12^0) & a_1 & \hfill 0.66 \cr 
(12^2) & e & \hfill -5.0\cr 
(21^1) & e & \hfill 4.07 \cr 
(30^0) & a_1 & \hfill -1.23 \cr} 
$$

\

$$
\matrix{ rms && 5.84 \,\, cm^{-1} \cr 
& \cr 
Parameters && \alpha = 3193.600 \hfill \cr 
&& \beta = 2507.157 \hfill \cr 
&& \gamma = 2807.833 \hfill \cr 
&& \cr 
&& \delta = -13.439 \hfill \cr 
&& \cr 
&& \alpha^{[2]} = -14.855 \hfill \cr 
&& \beta^{[2]} = -27.752 \hfill \cr 
&& \xi^{[2]} = - 28.043 \hfill \cr 
&& \cr
&& \epsilon = - 0.900 \hfill \cr } 
$$

\end